\documentstyle[prl,aps,twocolumn]{revtex}

\newcommand{\be}{\begin{equation}}
\newcommand{\ee}{\end{equation}}
\newcommand{\bea}{\begin{eqnarray}}
\newcommand{\eea}{\end{eqnarray}}

\newcommand{\Eq}{Eq. \ref}
\newcommand{\Fig}{Fig. \ref}

\def\<{\langle} 				
\def\>{\rangle} 				
\def\half{{1\over 2}}				

\begin{document}
\preprint{Submitted to {\em Physical Review Letters}}


\title{Gaussian resolutions for equilibrium 
density matrices}

\author{Pavel Frantsuzov$^{1,2}$, Arnold Neumaier$^3$ and 
Vladimir A. Mandelshtam$^1$}

\address{$^1$ Chemistry Department, 
University of California at Irvine, Irvine, CA 92697, USA}

\address{$^2$ Institute of Chemical Kinetics and Combustion RAS, 
Novosibirsk 630090, Russia}

\address{$^3$ Institut f\"ur Mathematik, Universit\"at Wien
Strudlhofgasse 4, A-1090 Wien, Austria;\\  
WWW: http://www.mat.univie.ac.at/$\sim$neum/}

\date{\today}

\maketitle

\abstract{ A Gaussian resolution method for the computation of
equilibrium density matrices $\widehat\rho_T$ for a general 
multidimensional quantum problem is presented. 
The variational principle applied to the ``imaginary time'' 
Schr\"odinger equation provides the 
equations of motion for Gaussians in a resolution of $\widehat\rho_T$
described by their width matrix, center and scale factor, 
all treated as dynamical variables.
The method is computationally very inexpensive, has 
favorable scaling with the system size and is
surprisingly accurate in a wide temperature range, 
even for cases involving quantum tunneling. 
Incorporation of symmetry constraints, 
such as reflection or particle statistics, is also discussed.
}} 




\bigskip
{\bf Propagating Gaussian resolutions.}
Computing the equilibrium 
quantum density matrix
\be\label{eq:rho}
\widehat \rho_T:=Z^{-1}e^{-\beta \widehat H},~~~
Z:=\mbox{Tr~} e^{-\beta \widehat H},
\ee
in dependence of the 
inverse temperature $\beta=1/kT$, 
or finding an equilibrium property of the type
\be\label{eq:Abeta}
\<\widehat A\>_T:=\mbox{Tr~}\widehat \rho_T\widehat A
\ee
for some operator $\widehat A$, is of great interest in 
statistical physics and molecular dynamics.
The most accurate methods, based on a spectral resolution
\[
\widehat \rho_T = \sum_n e^{-\beta E_n} |n\>\<n|
\]
are extremely expensive and limited to very few degrees of freedom.
Commonly used practical, albeit still very expensive, strategies
involve path integrals \cite{Feynman}.
An alternative semiclassical method was recently 
proposed \cite{makri2002},
but it is also expensive, while not very accurate.
In this letter we modify the Gaussian propagation techniques of solving
the time-dependent Scr\"odinger equation
\cite{Heller,Sawada,Coalson,Pattanayak} for the present case,
formally corresponding to propagation in imaginary time $\beta$,
and complement it by an important new idea: Monte Carlo-type sampling
is replaced by the propagation of a Gaussian resolution of 
$\widehat \rho_T$. This brings our semiclassical technique formally 
closer to the full quantum treatment by a spectral resolution.
It turns out to give a simple, very inexpensive and
surprisingly accurate alternative to the other approaches.
An extension to nonequilibrium problems is possible and will be 
discussed elsewhere.

Consider a $N$-body (or $D=3N$-dimensional) quantum system with 
Hamiltonian 
\be\label{eq:H}
\widehat H=\frac 1 2  \widehat p^{\rm T} M^{-1} \widehat p + U(x).
\ee
Here $\widehat p$ ($x$) define $D$-dimensional column vectors
of momentum (coordinate) operators and $M$ is the mass matrix.
The potential function $U(x)$ is assumed to be
representable as a sum of terms, each being 
a product of a complex Gaussian times a polynomial. 
(With this assumption a commonly used plane wave 
representation of $U(x)$ is covered.)

Given a grid $q_n$ spanning the physically relevant region in  
configuration space we put
\be \label{qbeta}
|q_{n,\tau}\>:=e^{-\tau \widehat H}|q_n\>,
\ee
with $\<x|q_n\>=\delta(x-q_n)$.
The resolution of identity can be approximated by
\be\label{eq:I}
\widehat I = \int \, d^{\rm D} q\,  | q\>\<   q|\approx 
\sum_n w_n \,| q_n\>\<   q_n|,
\ee
where the sum is over all the grid points and the 
$w_n$ are quadrature weights defined by the inverse local 
density of $q_n$. (For large number of dimensions $D$ 
a Monte Carlo procedure to generate $q_n$ may be adopted.)
Writing 
$e^{-\beta \widehat H}
=e^{-\beta \widehat H/2}\widehat Ie^{-\beta \widehat H/2}$
and inserting \Eq{eq:I}, we obtain 
\be\label{eq:Z}
Z\approx \sum w_n \<q_{n,\beta/2}| q_{n,\beta/2}\>,
\ee
\be\label{eq:rho2}
\widehat \rho_T\approx
Z^{-1}\sum w_n | q_{n,\beta/2}\>\<q_{n,\beta/2}|,
\ee
To find $|q_{n,\beta/2}\>$ we note that it
is the solution of the initial-value problem
\be \label{EqM}
\frac d {d\tau}|q_{n,\tau}\>=- \widehat H |q_{n,\tau}\>,~~~
|q_{n,0}\>=|q_n\>
\ee
We solve Eq.(\ref{EqM}) approximately using the ansatz
\cite{Heller,Sawada,Coalson,Pattanayak}
\be\label{eq:ansatz}
|q_{n,\tau}\>\approx|\lambda(\tau)\>
\ee
with 
\be\label{Gauss}
\<x|\lambda(\tau)\>
=\exp\left\{\gamma(\tau)-\half[x-q(\tau)]^T G(\tau) [x-q(\tau)]\right\}.
\ee
This is a Gaussian with center $q$, a real vector, 
width matrix $G$, a real symmetric and positive definite, and scale
$\gamma$, a real constant, and $\lambda:=(G,q,\gamma)$, 
a short-cut notation
containing all the Gaussian parameters. 

With this ansatz 
the maximum number of parameters corresponding to the use of full 
$D\times D$ dimensional matrix $G$ is  $(D+2)(D+1)/2$. 
This number may possibly be reduced assuming weak coupling 
between certain degrees of freedom and setting the corresponding 
matrix elements of $G$ to zero. The minimum number of $2D+1$ parameters
would correspond to using a diagonal width matrix $G$. 

To solve for $\lambda=\lambda(\tau)$
we follow the corresponding derivations 
\cite{Heller,Sawada,Coalson,Pattanayak} 
for the real-time dynamics 
of a Gaussian wavepacket by utilizing the variational 
principle 
\be\label{eq:var}
\left[{\partial L \over \partial\lambda'}\right]_{\lambda'=\lambda}=0
\ee
with the Lagrangian
\be\label{eq:L}
L=\left\< 
\lambda'(\tau) \left| {d\over d\tau}+\widehat H \right| \lambda(\tau) 
\right\>.
 \label{L}
\ee
Defining the two matrices
\be
K=\<  \lambda'|\lambda\>,~~~ H=\<  \lambda'| \widehat H|\lambda\>,
 \label{Hav}
\ee
\Eq{eq:var} can be rewritten as
\be
 \left[\frac{\partial^2 K}{\partial \lambda \partial\lambda'} 
\dot \lambda+\frac{\partial  H}{\partial \lambda'}\right]_
 {\lambda'=\lambda}=0,
 \label{EqMot}
\ee
thus providing the equations of motion for the Gaussian parameters 
$\lambda=\lambda(\tau)$.
The use of the Gaussian 
wavepackets and the assumed form of the Hamiltonian (\ref{eq:H}) 
allows one to evaluate the corresponding matrix elements and their 
derivatives in \Eq{EqMot} analytically.
(Efficient numerical expressions will be published elsewhere, but see 
refs.~\cite{Sawada,Pattanayak}).
Note that the commonly used
linearization of the potential \cite{Heller} reduces the complexity 
of the matrix elements evaluation, but also the accuracy, and is not 
used here.

The delta function in the initial condition of Eq.(\ref{EqM}) 
can be considered as a limit of the Gaussian (\ref{Gauss}) with 
infinite $G$. To avoid the singularities at $\tau=0$ we start 
instead at some small $\tau_0$ where we approximate $|q_{n,\tau_0}\>$ 
by a Gaussian that has finite width:
\bea\label{t0}
\langle x|\lambda(\tau_0)\rangle &\approx& 
\sqrt{\frac {\det M}{(2\pi\tau_0)^d}}\\\nonumber
&\times& \exp\left[-\frac 1 {2\tau_0} (x-q_n)^T M
(x-q_n)-U(q_n)\tau_0\right].
  \eea

To solve \Eq{EqMot}, we use the implicit integrator DASSL \cite{DASSL},
which has an error control and can be applied directly to \Eq{EqMot}.
(Probably, a standard numerical integrator could be utilized, if
\Eq{EqMot} is explicitly solved for $\dot \lambda$.)
Note that $\beta$ is small at large temperature $kT=1/\beta$, and 
only a few integration steps are needed. As $T$ decreases, the 
integration time becomes larger, and the variational approximation 
by Gaussians may become poorer.
We did not encounter any special numerical difficulties, except in 
the cases when the Gaussian width was too small.

Given (\ref{eq:ansatz}), we can rewrite (\ref{eq:rho2}) as
\be\label{eq:rho2a}
\widehat \rho_T\approx
Z^{-1}\sum w_n | \lambda_n(\beta/2)\>\<\lambda_n(\beta/2)|,
\ee
and get for expectations
\be\label{eq:Abeta'}
\<A\>_T\approx Z^{-1}
\sum w_n \<\lambda_n(\beta/2)| \widehat A | \lambda_n(\beta/2)\>.
\ee
The evaluation of $\<A\>$ requires the computation of many matrix 
elements of $\widehat A$ between new Gaussian wavepackets for every 
value of $T$. This can be done analytically if $A$ is polynomial
in $p,x$ or the product of a polynomial and a complex Gaussian.

\bigskip

{\bf Taking into account symmetry.}
Very often the system in question has symmetries, that is, 
the corresponding time-dependent
Schr\"odinger equation conserves certain symmetries. To show how to utilize 
this, we consider the example of reflection symmetry, 
$\widehat S \psi(x)=\psi(-x)$, and assume that
\be
|q_{n,\tau}^{\pm}\>
={1\over\sqrt{2}}(|q_{n,\tau}\>\pm \widehat S|q_{n,\tau}\>)
\ee
is a solution of \Eq{EqM} at any $\tau$.
Rewriting the resolution of identity,
\be\label{eq:I'}
\widehat I \approx 
\sum_n w_n (| q_n^+\>\<q^+_n|+| q_n^-\>\<q^-_n|),
\ee
we obtain
\bea\label{eq:Abeta''}
Z &\approx& 
\sum w_n (\<q_{n,\beta/2}^+| q_{n,\beta/2}^+\>
+\<q_{n,\beta/2}^-| q_{n,\beta/2}^-\>),
\\\nonumber \widehat \rho &\approx&
Z^{-1}\sum w_n (| q_{n,\beta/2}^+\>|\<q_{n,\beta/2}^+|
+| q_{n,\beta/2}^-\>|\<q_{n,\beta/2}^-|),
\\\nonumber
\<A\> &\approx& Z^{-1}
\sum w_n (\<q_{n,\beta/2}^+| \widehat A | q_{n,\beta/2}^+\> 
+\<q_{n,\beta/2}^-| \widehat A | q_{n,\beta/2}^-\>).
\eea

To approximate the solutions $|q_{n,\tau}^{\pm}\>$
we can replace the Gaussian in \Eq{eq:ansatz} by the symmetrized 
Gaussian:
\be
|\lambda_\pm\>={1\over\sqrt{2}}(|\lambda\>\pm \widehat S|\lambda\>),
\ee
where 
$$
\widehat S|G(\tau),q(\tau),\gamma(\tau)\> 
= |G(\tau),-q(\tau),\gamma(\tau)\>.
$$
Therefore, the matrix elements needed in \Eq{EqMot} can be evaluated
by taking the appropriate linear combinations of matrix elements
between single Gaussians.

\bigskip
{\bf Bose and Fermi statistics.}
Another important case of symmetry corresponds to 
the Bose or Fermi statistics. 
For a system of $N$ indistinguishable particles in
configuration space defined
by the coordinates
$$x=(x_1,\dots,x_N)^{\rm T}$$
we should symmetrize or antisymmetrize the wavepackets according to 
the particle statistics.
Let $\alpha=(\alpha_1,\dots,\alpha_N)$ define a permutation
of particle indices, and denote by $P_\alpha$ the permutation matrix
with 
\begin{figure}
\vspace{7cm}
\includegraphics{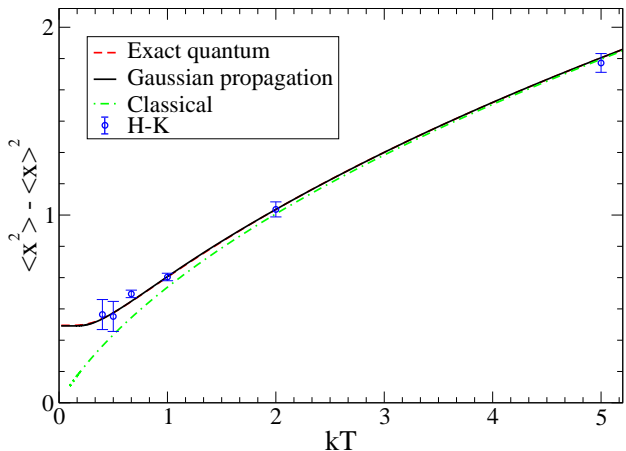}
\caption{\label{fig1}
Mean square displacement computed by four different methods for 
1D single well potential 
$U(x)=\half x^2 +0.1 x^4$. The semiclassical result labeled by 
``H-K'' is taken from ref.1 
The difference between 
the present and exact quantum result is not seen in the graph.
}
\end{figure}

\begin{figure}
\vspace{7cm}
\includegraphics{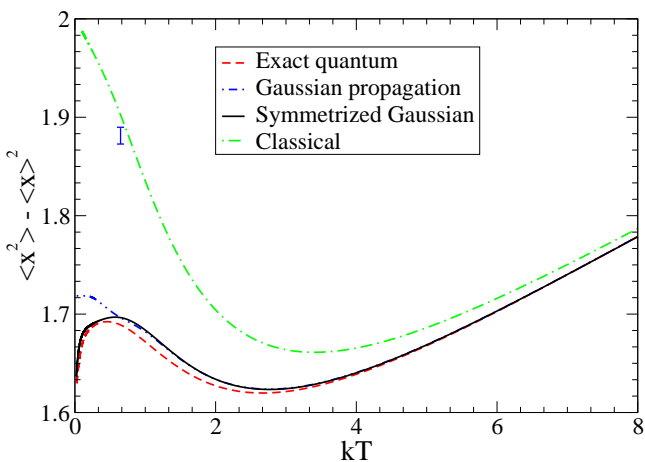}
\caption{\label{fig2} The mean square displacement for 
the symmetric double-well 
potential $U(x)=4-4 x^2 + x^4$ computed by four different methods.
}
\end{figure}
$$P_\alpha x = (x_{\alpha_1},\dots,x_{\alpha_N})^{\rm T}.$$
A symmetrized wavepacket $|\lambda_\pm\>$ can be defined by
\be \label{Symm}
\<  x|\lambda_\pm\>=
\sum_{\alpha}\mbox{sign}_\pm(\alpha)\<P_\alpha x|\lambda\>.
\ee
Here boson statistics has $\mbox{sign}_-(\alpha)=1$; in the case of 
fermion statistics, $\mbox{sign}_+(\alpha)=1$ for even permutations
and  $\mbox{sign}_+(\alpha)=-1$ for odd permutations.
The equations of motion (\ref{EqMot}) have the same general form.
Since \Eq{Symm} implies
\[
|\lambda_\pm\>=\sum_{\alpha}\mbox{sign}_\pm(\alpha)P_\alpha^*|\lambda\>
\]
and one easily verifies
\[
P_\alpha^*|G,q,\gamma\>=|P_\alpha^*GP_\alpha,P_\alpha^*q,\gamma\>,
\]
here again, matrix elements with symmetrized Gaussians 
$|\lambda_\pm\>$ are simple linear combinations of suitable matrix 
elements with unsymmetrized Gaussians. Similar considerations apply 
to multiparticle systems with few indistinguishable particles, 
where the sum(\ref{Symm}) has only a few terms and the calculations 
remain feasible.

\bigskip
\noindent 
{\bf Numerical examples}

\bigskip

\begin{figure}
\vspace*{7cm}
\includegraphics{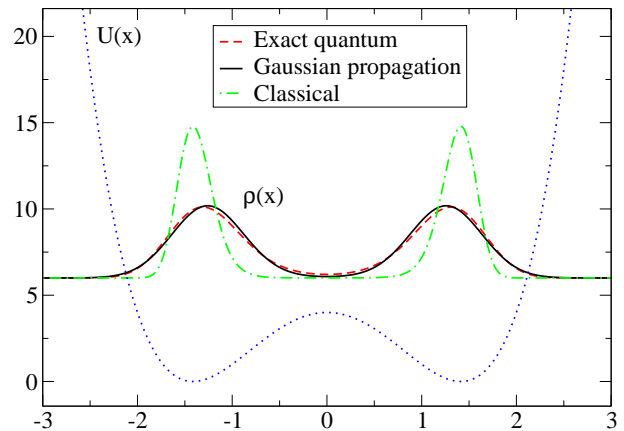}
\caption{\label{fig3}
The density $\rho(x)=\<x|\widehat\rho|x\>$
at $kT=0.5$ computed by three different methods for the symmetric
double-well
potential $U(x)$ as in \Fig{fig2}.
}
\end{figure}

{\bf 1D single-well problem.}
We first apply the method to a 1D problem of ref.~\cite{makri2002}
with potential $U(x)=\half x^2 +0.1 x^4$ to compute the 
mean square displacement $\<x^2\>-\<x\>^2$ as a function of 
temperature $T$. 
A grid of 10 equidistant points was taken in the interval $-5<q_n<5$.
(Here and in all the following examples the reported 
results fully converged with respect to the grid size.)
Each of the 10 Gaussians was then propagated by 
\Eq{EqMot} starting with $\tau_0=0.01$ up to $\tau=50$. 
The result obtained by the present method 
is hardly distinguishable from the converged quantum calculation
using diagonalization of the
Hamiltonian in a large basis. 
In \Fig{fig1} these results are also compared to the classical 
Boltzmann average and to the semiclassical calculation using the 
Herman-Kluk propagator \cite{makri2002}.
Note that in the latter case a much more expensive Monte Carlo method 
was used with $10^5$ classical trajectories,
still resulting in relatively big statistical errors.

\begin{figure}
\vspace*{13.2cm}
\includegraphics{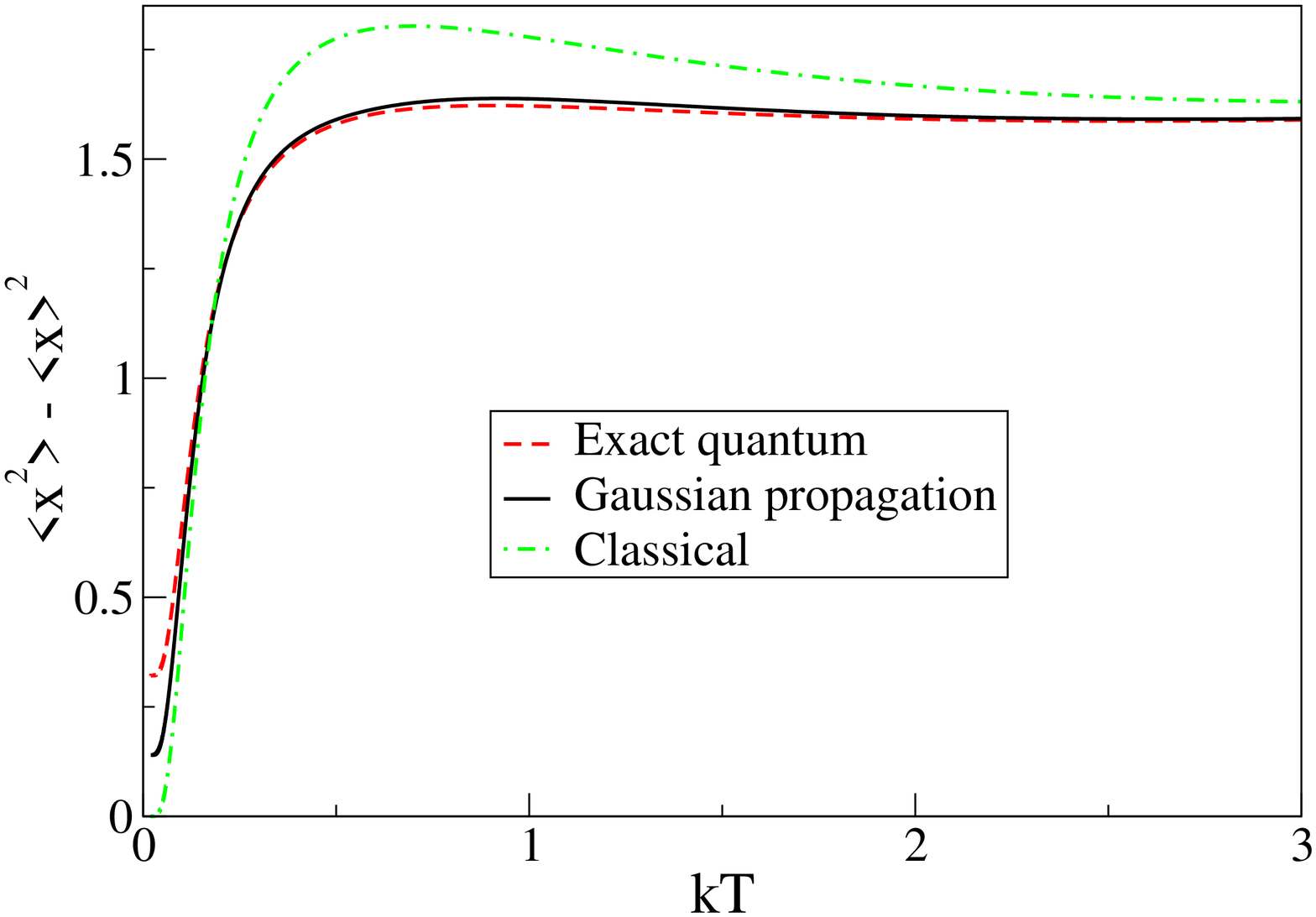}
\includegraphics{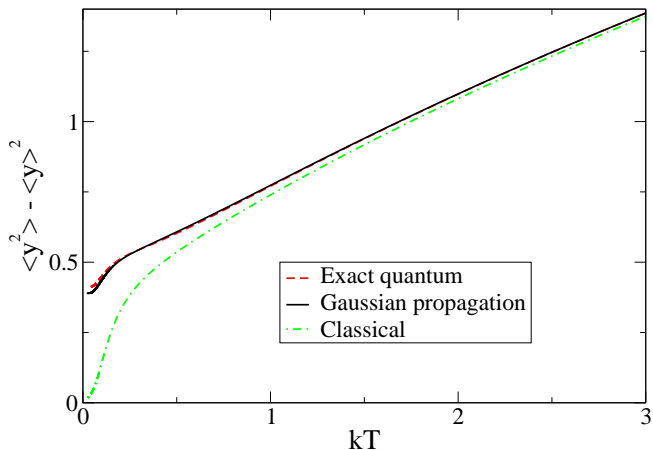}
\caption{\label{fig4} The mean square displacements computed 
by three different methods for the asymmetric 2D double-well potential
$U(x,y)=(x^2-2)^2+0.1x+0.5(y-0.5x)^2+0.1y^4$
}
\end{figure}

We note that a similar high accuracy was achieved for a 2D single 
well problem (not shown here).

\bigskip 
{\bf 1D symmetric double-well problem.}
Here the method was applied to a problem with the potential 
$U(x)=4-4 x^2 + x^4$.
Now a grid of 14 points with $-3<q_n<3$ was used.
The corresponding results using both the unsymmetrized and symmetrized 
gaussians are shown in \Fig{fig2}.
The agreement between the exact result 
(fully converged diagonalization of $\widehat H$ in a large basis) 
and that computed by the 
present method is unexpectedly excellent even at 
quite low temperatures, well below the potential barrier. 
For the unsymmetrized case a 
significant deviation from the exact (and symmetrized) result
occurs only at low temperatures where the small 
tunneling splitting causes the quantum
observables change rapidly with $T$.

In \Fig{fig3} we show the density profile for 
the same system computed using the same set of 14 symmetrized 
Gaussians at $kT=0.5$ together with the exact quantum 
and classical results.

\bigskip 
{\bf 2D asymmetric double-well problem.}
To further demonstrate the method we apply it to a 2D problem with 
asymmetric double-well potential 
$U(x,y)=(x^2-2)^2+0.1x+0.5(y-0.5x)^2+0.1y^4$. An equidistant 
$16\times 16$ grid $q_n$ 
in a square box $[-4;4]\times [-4;4]$ was used. The results 
for the mean square displacements shown in \Fig{fig4}   
are again surprisingly accurate, except for very low temperatures.

%
%

\bigskip

\noindent
{\bf Acknowledgement.}\hskip 0.2in
V.A.M. acknowledges the NSF support, grant CHE-0108823. He is
an Alfred P. Sloan research fellow.


\end{document}